\documentclass[a4paper,11pt]{amsart}
\pdfoutput=1
\usepackage{amsfonts,amsthm,amsmath,amssymb,bbold, upgreek, dsfont, tipa, epsfig, color, inconsolata, framed, mathtools, sidecap, graphics, graphicx, wrapfig, overpic}

\usepackage[utf8]{inputenc}
\usepackage[font=small,labelfont=bf]{caption, subcaption}
\usepackage[T1]{fontenc}
\usepackage[ruled,vlined, norelsize]{algorithm2e}

\title{From birds to bacteria: generalised velocity jump processes with resting states.}

\date{\today}

\author{Jake P. Taylor-King}
\dedicatory{Contact email: \texttt{jake.taylor-king@sjc.ox.ac.uk}.}
\author{Emiel van Loon}
\author{Gabriel Rosser}
\author{S. Jon Chapman}

\theoremstyle{plain}

\theoremstyle{definition}

\theoremstyle{remark}

\numberwithin{equation}{section}

\newcommand{\vect}[1]{\boldsymbol{#1}}
\newcommand{\vectornorm}[1]{\left|\left|#1\right|\right|}

\newcommand{\p}{\partial}
\newcommand{\dd}{\text{d}}

\begin{document}

\begin{abstract}
There are various cases of animal movement where behaviour broadly switches between two modes of operation, corresponding to a long distance movement state and a resting or local movement state. Here a mathematical description of this process is formulated, adapted from Friedrich \emph{et. al.} \cite{Friedrich_2006}. The approach allows the specification \emph{any} running or waiting time distribution along with any angular and speed distributions. The resulting system of partial integro-differential equations are tumultuous and therefore it is necessary to both simplify and derive summary statistics. An expression for the mean squared displacement is derived which shows good agreement with experimental data from the bacterium \emph{Escherichia coli} and the gull \emph{Larus fuscus}. Finally a large time diffusive approximation is considered via a Cattaneo approximation \cite{Hillen_2004}. This leads to the novel result that the effective diffusion constant is dependent on the mean and variance of the running time distribution but only on the mean of the waiting time distribution.
\end{abstract}
\maketitle

\section{Introduction.}
In nature, organisms whose sizes differ by many orders of magnitude have been observed to switch between different modes of movement. For instance, the bacterium \emph{Escherichia coli} changes the orientation of one or more of its flagella between clockwise and anticlockwise to achieve a \emph{run-and-tumble} like motion \cite{Berg_1983, Berg_1990}. As a result, during the runs, we see migration-like movement and during the tumbles, we see resting or local diffusion behaviour\footnote{The reason for observing diffusive-like behaviour is due to the bacterium's small size - which is on the length-scale of micrometres.}. To add to this complexity, it should be noted that the direction of successive runs are correlated. On a larger scale one could consider migratory movements of vertebrates where individuals often travel large distances intermittent with stop-overs to rest or forage. An example, used in this paper, is the lesser black-backed gull (\emph{Larus fuscus}). Individuals of this species that breed in the Netherlands migrate southwards during Autumn. Even though the scales involved in these two processes differ by many orders of magnitude, one can use a similar mathematical framework to model the observed motion.

The use of mathematical models to describe the motion of a variety of biological organisms, including bumblebees \cite{kareiva83}, plants \cite{cain90} and zebra \cite{brooks08} has been the subject of much research interest for several decades.  Early approaches were predominantly centred on the position jump model of motion \cite{Brenner_1998,skellam51}, where agents instantaneously change position according to a distribution kernel and are interspersed with waiting periods of stochastic length. 
The position jump framework suffers from the limitation that correlations in the direction of successive runs are difficult to capture, this correlation however is present in many types of movement \cite{Marsh_1988}.  Furthermore, the diffusive nature of the position jump framework results in an unbounded distribution of movement speeds between successive steps.  A related framework that is arguably more realistic for modelling the motion of organisms is the velocity jump (VJ) model \cite{Othmer_1988}, in which organisms travel with a randomly-distributed speed and angle for a finite duration before undergoing a stochastic reorientation event.

In most formulations of the velocity jump process, there is an assumption that events occur as a Poisson process, which is manifested as a constant rate parameter in the resulting differential equation. In the position jump framework, non-exponentially distributed wait times and non-Gaussian kernel processes have been formulated, although this led to fractional diffusion equations \cite{Klafter_1987, Metzler_2000}. Recently, it has become clear how to extend the velocity jump framework to allow for more general distributions of interest \cite{Friedrich_2006_2, Friedrich_2006}.

In many velocity jump models, it is assumed that resting states are largely negligible \cite{Erban_2004, Erban_2005}, this can be attributed to a focus on organisms with only momentary resting states, this has the benefit of alleviating some mathematical complexity whilst not changing the result significantly \cite{Erban_2004}. However, in the work by Othmer \cite{Othmer_1988} and Erban \cite{Erban_2007}, it was shown that resting states can be included and are sometimes required in order to obtain adequate fits to experimental data \cite{Othmer_1988}. Our goal in this paper is to extend the work by Friedrich \emph{et. al.} \cite{Friedrich_2006} to allow for resting states - which are non-negligible - following the methodology of Othmer \cite{Othmer_1988}. 

The mathematical complexity of Friedrich's model is such that finding solutions analytically or numerically is, in general, impractical. In the original paper, simplifications were made which led to a fractional Kramers-Fokker-Planck equation, which has a known analytic solution \cite{Friedrich_2006}. However, the simplifications relevant to a physical system are seldom relevant to a biological one. For instance, the original formulation related to non-Gaussian kinetics in a weakly damped system; however, we are considering self-propelled particle models where biological agents generate their own momentum. In the absence of such obvious simplifications for our system, we instead exploit methods to extract summary statistics from the governing equations, which may in turn be compared with experimental data.

After presenting the model of interest we derive the mean squared displacement (MSD). As we have high-quality data available relating to the movement of \emph{E. coli} and \emph{L. fuscus}, we show that the MSD for the model and experimental data align. What is novel about our approach is that, provided the two discrete modes of operation constitute a good model, the parameters can be extracted on a microscopic scale prior to any numerical solution and then macroscopic behaviour can be derived \emph{without} optimising or trying to fit data \emph{a posteriori}. 

Since the dynamics of the experimental data and those of the generalised velocity jump model achieve a close match, we explore numerically tractable simplifications to the equations of interest. Most notably, we investigate the Cattaneo approximation, following the work by Hillen \cite{Hillen_2003, Hillen_2004}. 

Finally, it should be noted that the model presented does not take into account interactions between biological agents or even interactions with the environment. Whilst such effects are beyond the scope of the current study, it should be possible to extend the theory to incorporate these phenomena.  In particular, the velocity jump process has roots in Kinetic Theory and as such, similar to how atoms attract and repel one another, models have been developed for biological agents to act comparably \cite{Carillo_2009, Degond_2004, Naldi_2010}. Equally, there is similar work detailing interactions between a biological agent and its environment, both fixed environments and signalling via diffusing chemical gradients \cite{Chauviere_2010, Erban_2004, Erban_2005}.

\section{Two-state generalised velocity jump process.}\label{TwoStateGVJ}
Consider a biological agent that switches stochastically between running and resting behaviour. During a running phase, the organism travels with constant velocity; during a resting phase, it remains stationary. Upon resuming a run following a rest, a new velocity is selected randomly. This motion is governed by three primary stochastic effects. We specify these by probability density functions (pdfs), as given below.
\begin{itemize}
\item[i.)] Waiting time: The time spent during a resting phase, denoted $\omega$ is governed by the pdf $f_{\omega}(t)$, where $\int_0^\infty f_\omega (t)\dd t=1$.
\item[ii.)] Running time: The time spent during a running phase, denoted $\tau$ is governed by the pdf $f_{\tau}(t)$, where $\int_0^\infty f_\tau (t)\dd t=1$.
\item[iii.)] Reorientation: We allow velocities from one run to another to be correlated from before to after a rest. Suppose the previous running phase (pre-rest) had an associated velocity $\vect{v}'\in V$, for velocity space $V\subset \mathds{R}^n$ in $n$ spatial dimensions, then we write the new (post-rest) velocity as $\vect{v}\in V$ which is newly selected upon entering a running phase. The selection of $\vect{v}$ is dependent on $\vect{v}'$ and is governed by the joint pdf $T(\vect{v},\vect{v}')$. 

We assume that this reorientation pdf is separable, so that $T(\vect{v},\vect{v}') = g(\vect{\theta}, \vect{\theta}')h(s, s')/s^{n-1}$ where $\vect{\theta}$ is a vector of length $(n-1)$ containing angles and $s = \vectornorm{\vect{v}}$ is the speed. In two dimensions, the turning kernel is decomposed as follows.
\begin{quotation}
\item[a.)]  The angle distribution: $g(\theta, \theta')$, requires the normalisation $\int_0^{2\pi}g(\theta, \theta')\dd \theta = 1$.
\item[b.)] The speed distribution: $h(s, s')$, requires the normalisation $\int_0^{\infty}h(s, s')\dd s = 1$.
\end{quotation}
\end{itemize}

To further reinforce the process we are describing, we give a simple Gillespie algorithm \cite{Gillespie_1977} for generating a sample path up until time $T_{\text{end}} > 0$. It should be noted that the sample path will need to be truncated as it will generate positions past the end time.

\begin{algorithm}[H]
\DontPrintSemicolon
 \KwData{Initialise time $t=0$, starting position at $\vect{x}(t=0) = \vect{x}_0$ and starting velocity at $\vect{v}(t=0) =\vect{v}_0$.}
 Choose state of particle, for instance, assume particle has just initiated a running state. \;
 \While{$t < T_{\text{end}}$}{
  Sample time spent running $\tau\sim f_{\tau}(t)$.\;
  Update position: $\vect{x}(t+\tau) \leftarrow \vect{x}(t) + \tau\vect{v}(t)$.\;
  Sample time spent waiting $\omega\sim f_{\omega}(t)$.\;
  Update position: $\vect{x}(t+\tau+\omega) \leftarrow \vect{x}(t+\tau)$.\;
  Sample new velocity for next running phase: $\vect{v}(t+\tau+\omega)\sim T(\cdot, \vect{v}(t))$.\;
  Update time $t \leftarrow t + \tau + \omega$.\;
  }
 \caption{Algorithm to generate a single generalised velocity jump sample path.}
\end{algorithm}
By  considering the density of particles in a running state and the density of particles in a resting state, we can write down coupled differential equations for these states. We define $p=p(t,\vect{x},\vect{v})$ to be the density of particles at position $\vect{x}\in\Omega\subset\mathds{R}^n$, with velocity $\vect{v}\in V\subset\mathds{R}^n$ at time $t\in \mathds{R}^+$ and $r = r(t,\vect{x},\vect{v})$, the density of those particles resting at $(t, \vect{x})\in\mathds{R}^+\times \Omega$, having just finished a jump of velocity $\vect{v}\in V$. Note that this encodes an orientation to the resting state.

The derivation for this two-state generalised velocity jump process through the use of Laplace transforms is provided in Appendix \ref{AppA}. Our analysis leads to the following equations
\begin{eqnarray}\label{p_forward}
\left(\frac{\p}{\p t} + \vect{v}\cdot{\nabla_{\vect{x}}} \right) p(t,\vect{x},\vect{v}) = -\int_0^t \Phi_\tau(t-s) p(s,\vect{x} - (t-s)\vect{v},\vect{v}) \dd s \\ + \int_0^t \Phi_{\omega}(t-s)\int_V T(\vect{v},\vect{v}')r(s,\vect{x},\vect{v}')\dd\vect{v}'\dd s, \nonumber
\end{eqnarray}
and
\begin{equation}\label{r_forward}
\frac{\p}{\p t}r(t,\vect{x},\vect{v}) = -\int_0^t \Phi_\omega(t-s) r(s,\vect{x},\vect{v}) \dd s + \int_0^t \Phi_{\tau}(t-s)p(s,\vect{x}-(t-s)\vect{v},\vect{v}) \dd s,
\end{equation}
where the delay kernels, $\Phi_i$ for $i = \tau, \omega$, are defined in Laplace space by
\begin{equation}\label{f_psi_conversion}
\bar{\Phi}_i (\lambda) = \frac{\lambda \bar{f}_i (\lambda)}{1 - \bar{f}_i(\lambda)} \quad \text{for }i=\tau , \omega .
\end{equation}
where $\bar{f}_i$ is the Laplace transform of the pdf for the running and waiting time respectively. When the waiting time is chosen as exponential\footnote{This can be seen simply for exponential distribution with mean $\chi^{-1}$,
\begin{equation}
f_i(t) = \chi_i e^{-\chi_i t}, \quad\implies\quad \bar{f}_i(\lambda) = \frac{\chi_i}{\lambda + \chi_i} ,\quad\implies\quad \bar{\Phi}_i (\lambda) = \chi_i ,\quad\implies\quad \Phi_i (t) = \chi_i \delta (t),  \quad \text{for }i=\tau , \omega ,
\end{equation}
where $\delta$ is the Dirac delta function.}, this is consistent with work by Othmer \cite{Othmer_1988} and Rosser \cite{Rosser_2012}. 

Finding closed forms of $\Phi_i(t)$ is non-trivial for most choices of distribution $f_i(t)$. In Appendix \ref{AppB}, we examine the small time behaviour of $\Phi$ and identify the sizes of potential impulses at $t=0$. For the remaining non-singular behaviour, in the cases where we know the Laplace transform of $f_i(t)$, we then have an analytic expression for $\bar\Phi (\lambda)$, which can be inverted numerically using either a Talbot inversion or an Euler inversion \cite{Abate_1995, Murli_1990}.

\section{Mean-Squared Displacement.}
Equations (\ref{p_forward}--\ref{r_forward}) give us a system of delay-integro-partial differential equations with $(2n + 1)$ degrees of freedom. With this level of complexity, a full analytic or numerical solution is impractical without first making simplifications. We therefore first consider how to estimate the second spatial moment, i.e. the mean squared displacement \cite{Othmer_1988}.

For the test function $\varphi = \varphi(\vect{x}, \vect{v})$, we consider for arbitrary density $\rho = \rho(t,\vect{x},\vect{v})$, 
\begin{equation}
Q_\rho(\varphi,t) = \int_V \int_\Omega \varphi (\vect{x}, \vect{v})\rho(t,\vect{x},\vect{v}) \dd \vect{x} \dd \vect{v}.
\end{equation}
This gives the expected value of $\varphi$ over the space $V\times\Omega$ at time t, weighted by density $\rho$. By using test functions $\varphi = 1, \vectornorm{\vect{x}}^2, \vect{v}\cdot\vect{x}, \vectornorm{\vect{v}}^2$, we associate $N_\rho(t) = Q_\rho(1,t)$ as the number of particles in state $\rho$ and then $D_\rho^2 (t)= Q_\rho (\vectornorm{\vect{x}}^2, t)$, $B_\rho(t) = Q_\rho (\vect{v}\cdot\vect{x}, t)$ and $V_\rho^2 (t)= Q_\rho (\vectornorm{\vect{v}}^2, t)$ as the mean squared displacement, the mean velocity-displacement and the mean squared velocity weighted by $\rho$, respectively. We can then obtain a closed system of integro-differential equations for these quantities.

It first requires however, that we make some assumptions on the Turning kernel $T$. By considering that the mean post-turn velocity has the same orientation as the previous velocity, we define the index of persistence $\psi_d$ via the relation
\begin{equation}
\bar{\vect{v}}(\vect{v}') = \int_V \vect{v}T(\vect{v},\vect{v}') \dd \vect{v} = \psi_d \vect{v}'.
\end{equation}
Informally, this means that turning angles between consecutive velocities have zero mean. We also require that the average mean squared speed is a constant  
\begin{eqnarray}
S_T^2(\vect{v}') = S_T^2 = \int_V \vectornorm{\vect{v}}^2 T(\vect{v},\vect{v}') \dd \vect{v} ,
\end{eqnarray}
this corresponds to a memoryless turning kernel in speed, i.e. $h(s,s') = h(s)$.
Finally, for unconstrained motion where $\Omega = \mathds{R}^n$, we see that delays in space correspond to inclusion of other moments, i.e.
\begin{eqnarray}
\int_V \int_\Omega \vectornorm{\vect{x}}^2 \rho(t,\vect{x}-c\vect{v},\vect{v}) \dd \vect{x} \dd \vect{v} &=& \int_V \int_\Omega \vectornorm{\vect{x} + c\vect{v}}^2 \rho(t,\vect{x},\vect{v}) \dd \vect{x} \dd \vect{v} ,\\ &=& \int_V \int_\Omega \left(\vectornorm{\vect{x}}^2 + 2c(\vect{v}\cdot\vect{x}) + c^2 \vectornorm{\vect{v}}^2 \right)\rho(t,\vect{x},\vect{v}) \dd \vect{x} \dd \vect{v} , \\ &=&   D_\rho^2(t) + 2c B_\rho(t) + c^2 V_\rho^2(t) ,
\end{eqnarray}
and similarly
\begin{eqnarray}
\int_V \int_\Omega (\vect{v}\cdot \vect{x}) \rho(t,\vect{x}-c\vect{v},\vect{v}) \dd \vect{x} \dd \vect{v} &=& \int_V \int_\Omega  (\vect{v}\cdot \vect{x} + c\vect{v}) \rho(t,\vect{x},\vect{v}) \dd \vect{x} \dd \vect{v} , \\ &=&   B_\rho(t) + c V_\rho^2(t) .
\end{eqnarray}
For conservation of mass, i.e. $N_p(t) + N_r(t) = N_0$, we see that
\begin{eqnarray}\label{MSD_first_eq}
\frac{\dd N_p(t)}{\dd t} = - \frac{\dd N_r(t)}{\dd t} = -\int_0^t \Phi_\tau (t-s)N_p(s)\dd s + \int_0^t \Phi_\omega (t-s)N_r (s) \dd s.
\end{eqnarray}
Equally, we obtain a system of equations for the mean squared displacement
\begin{eqnarray}
\frac{\dd D_p^2(t)}{\dd t} - 2B_p(t) = -\int_0^t \Phi_\tau(t-s)\left[ D_p^2(s) + 2(t-s)B_p(s) + (t-s)^2V_p^2(s)\right]\dd s, \\ + \int_0^t \Phi_\omega(t-s) D_r^2(s) \dd s = - \frac{\dd D_r^2(t)}{\dd t}. \nonumber
\end{eqnarray}
For the mean velocity-displacement, we see that
\begin{eqnarray}
\frac{\dd B_p(t)}{\dd t} = V_p^2(t) - \int_0^t \Phi_\tau (t-s) \left[ B_p(s) + (t-s)V_p^2(s)\right] \dd s, \\ + \psi_d \int_0^t \Phi_\omega (t-s) B_r(s) \dd s, \nonumber
\end{eqnarray}
and
\begin{equation}
\frac{\dd B_r(t) }{\dd t} = - \int_0^t \Phi_\omega (t-s) B_r(s) \dd s+ \int_0^t \Phi_\tau (t-s) \left[ B_p(s) + (t-s)V_p^2(s)\right] \dd s. 
\end{equation}
Finally, for the second velocity moment:
\begin{eqnarray}
\frac{\dd V_p^2(t)}{\dd t} = - \int_0^t \Phi_\tau (t-s) V_p^2(s) \dd s +S_T^2 \int_0^t \Phi_\omega (t-s) N_r(s) \dd s, \\ \frac{\dd V_r^2(t)}{\dd t} = - \int_0^t \Phi_\omega (t-s) V_r^2(s) \dd s+ \int_0^t \Phi_\tau (t-s) V_p^2 (s) \dd s. \label{MSD_last_eq}
\end{eqnarray}
Equations (\ref{MSD_first_eq})--(\ref{MSD_last_eq}) above correspond to a system of 8 equations, or 7 unique equations once we impose conservation of mass. In the next section, we solve these equations numerically, the integrals are calculated using the trapezium rule along with a Crank-Nicholson scheme for the remaining differential operators, both of these methods are second-order accurate.
 
\section{Comparison between Theory and Experiment.}\label{CompExpTheo}
In this study, we consider experimental data relating to the bacterium \emph{E. coli} and the lesser black-backed gull \emph{L. fuscus}. Both of these exhibit somewhat similar behaviour, however at scales many orders of magnitude apart.
\subsection{E. coli.}
There is a large collection of work relating to studying the run-and-tumble motion as exhibited in many flagellated bacteria \cite{Berg_1972,frymier95,Rosser_2013,wu06}. A case of particular interest to many is \emph{E. coli}, perhaps due to the fact that its internal signalling pathways are less complex than those of other chemotactic bacteria \cite{porter08}. Most available literature points to both the running and resting times being exponentially distributed \cite{Berg_1990}. This exponential parameter can change as a response to its environment and has led to a multitude of papers showing that this mechanism leads to chemotaxis either towards nutrients or away from toxins \cite{Erban_2004, Erban_2005}.

\begin{wrapfigure}{r}{0.4\textwidth}
  \begin{overpic}[width=0.4\textwidth]{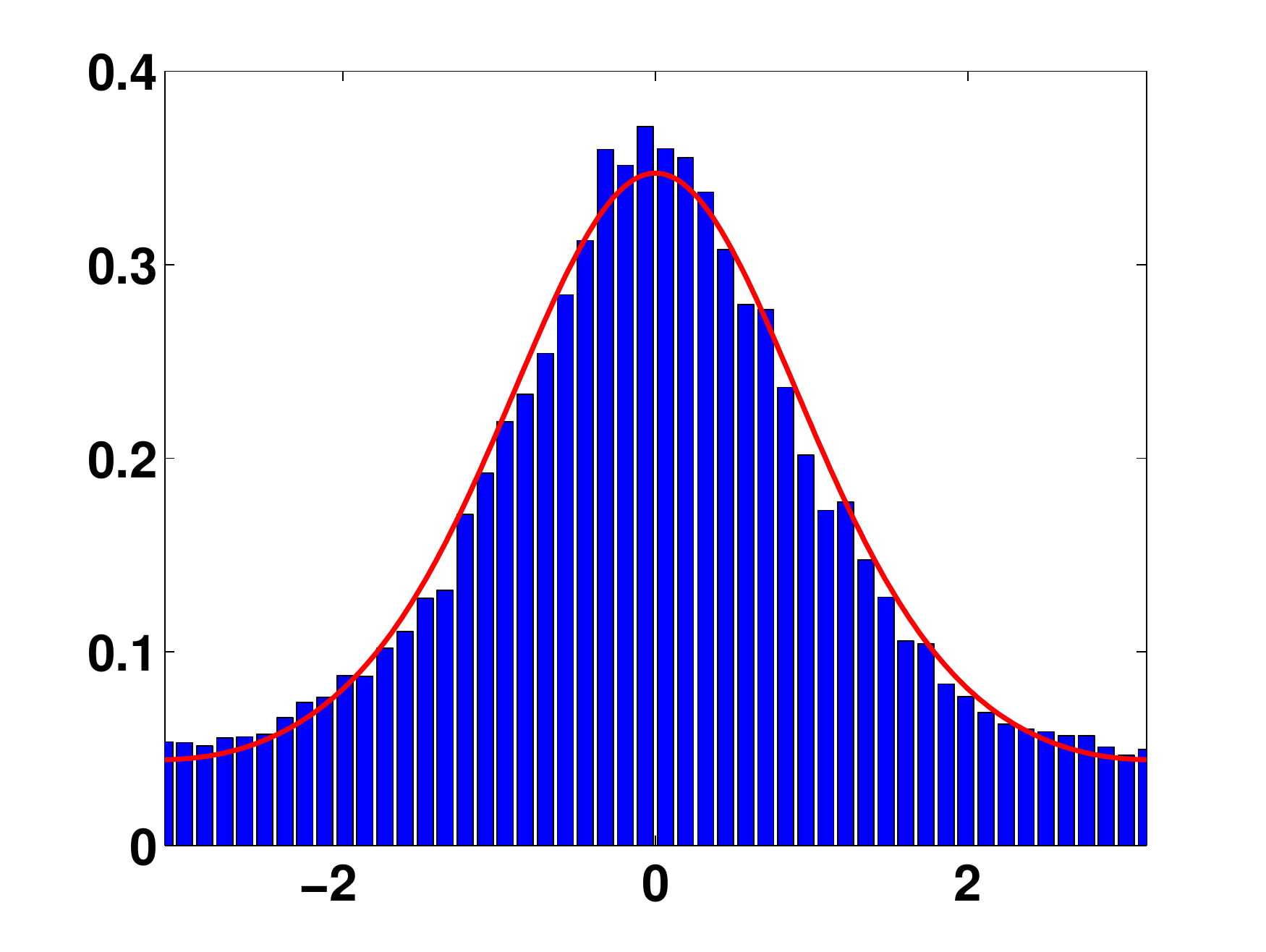}
  	\put(1,24){\rotatebox{90}{\small{Prob. Density}}}
	\put(33,0){\small{Angle in radians}}  	  
  \end{overpic}
   \caption{\footnotesize{Fit between experimentally observed values of angle changes from run-to-run \emph{(blue)} and the probability density function for the Von-Mises distribution \emph{(red)}.}}
   \label{VonMises_Ecoli}
\end{wrapfigure}
In our case however, we do not consider \emph{E. coli} in any chemical gradient but just swimming freely. The dataset used here has previously been described in studies by Rosser \emph{et. al.} \cite{Rosser_2014, Rosser_2013}. In brief, the data was obtained by performing video microscopy on samples of free-swimming \emph{E. coli}, from which tracks were extracted using a kernel-based filter \cite{Wood_2012}.  The tracks were subsequently analysed using a Hidden Markov Model to infer the state (running or resting) attributed to the motion between each pair of observations in a track \cite{Rosser_2013}.  From the annotated tracks, it is possible to extract the angle changes observed between running phases and parameters for the exponential running and waiting pdfs along with speed distributions.

\subsubsection{Results.}
In Figure \ref{VonMises_Ecoli}, we see that from run-to-run, the distribution of angles is approximately a wrapped normal distribution. For mathematical ease, consider the Von Mises distribution as an approximation as plotted in red and given by the probability density function
\begin{equation}\label{eq_vm_dist}
\Theta (\theta | \mu, \kappa) = \frac{e^{\kappa \cos (\theta - \mu)}}{2\pi I_0(\kappa)}, \quad\text{for }\kappa > 0, \mu\in\mathds{R},
\end{equation}
where $I_0(\cdot)$ is the modified Bessel function of order zero. By assuming $g(\theta, \theta') = \Theta (\theta - \theta')$, i.e. symmetry around the previous direction, we can specify $\mu = 0$, and find $\kappa$ through maximum likelihood estimation. It has been shown that for the choice of a Von Mises distribution, in two dimensions ($n=2$), the index of persistence is given by $\psi_d = I_1(\kappa) / I_0 (\kappa)$.
\begin{figure}[h!]
  \begin{overpic}[width=0.5\textwidth]{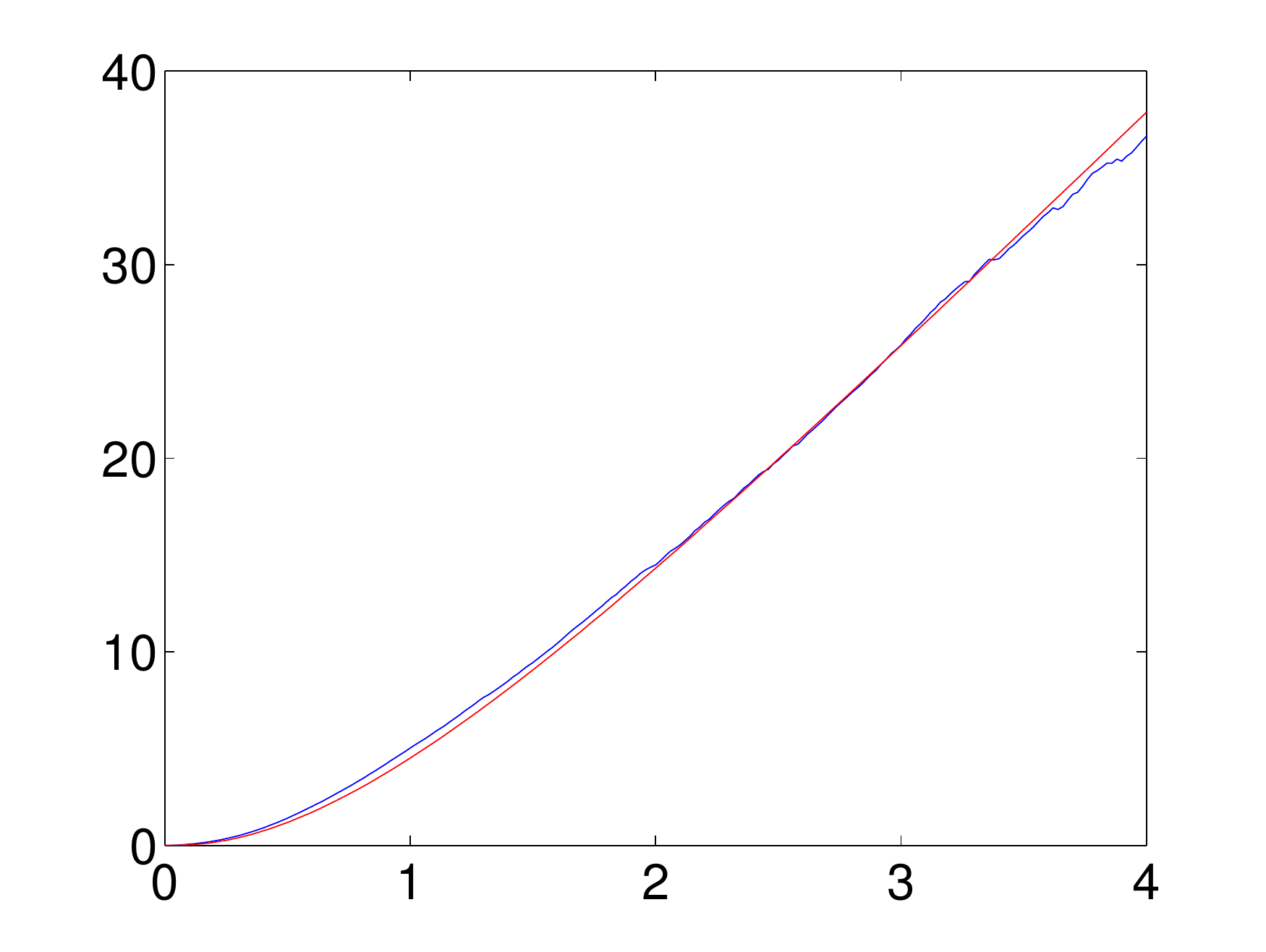}
  	\put(3,10){\rotatebox{90}{Mean sq. displacement $(\mu \text{m})^2$}}
	\put(44,0){Time (s)}  
  \end{overpic}  
  \caption{\footnotesize{{Comparison between system of equations (\ref{MSD_first_eq}--\ref{MSD_last_eq}) and \emph{E. coli} data. In red, the theoretical value of $(D_p^2 + D_r^2)/N_0$ is plotted and in blue there is the experimentally derived average MSD calculated from the bacterium's initial position. From the data, we determined that $\tau \sim \text{Exp}(2.30)$, $\omega \sim \text{Exp}(11.98)$. Equally, for the system of differential equations, we specify $N_p(0) = 66$, $N_r(0) = 1802$, $\psi_d = 0.46$ and $S_T^2 = 9.26\,(\mu\text{m})^2/\text{s}$. The initial state for all other differential equations is set to zero, except for $V_p^2(0) = S_T^2 N_p^2(0)$.}}}
  \label{MSD_eColi}
\end{figure}

It should be noted that from the literature, \emph{E. coli} is thought to have a bi-modal distribution around the previous direction \cite{Berg_1972}, the validity of this is hard to confirm as previous data was hand annotated and it is hard to specify the state of the bacterium when diffusion effects are also in place. Whilst we had more data available to us and used automated tracking methods, it could well be that our method heavily biases walks towards normally distributed reorientation.

Through the HMM technique as outlined in \cite{Rosser_2014, Rosser_2013}, estimates for the exponential parameters were found to be $\tau \sim \text{Exp}(2.30)$ and $\omega \sim \text{Exp}(11.98)$. The mean squared speed whilst running was also calculated to be $S_T^2 = 9.26\,(\mu\text{m})^2/\text{s}$. In Figure \ref{MSD_eColi}, we plot the mean squared displacement over time. We clearly see that over the average of 1868 paths, we get a very good match between theory and experiment. We note that the videos were taken from a fixed position, where bacteria would swim in and out of the shot. By considering the average speeds of \emph{E. coli} along with the size of the viewing window, one can stipulate that by only considering the MSD before 4 seconds, we can achieve a good estimate. Note that we lose a small amount of data over time as bacterium swim out of the observation window, at later times this ruins the validity of the MSD curve.

\subsection{Lesser black-backed gull.}

\begin{wrapfigure}{l}{0.4\textwidth}
  \begin{center}
	\includegraphics[natwidth=400, natheight=350, width=0.4\textwidth]{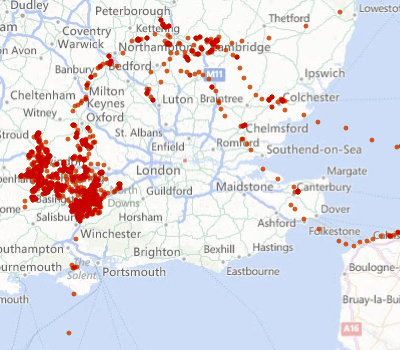}
  \end{center}
  \caption{\footnotesize{Sample path from bird data.}}\label{bird_sample_path_map}
\end{wrapfigure}

In this section we consider Lesser black-backed gulls that breed on Texel (the Netherlands). During their non-breeding period (August to April), these birds interchange between localised movements (or resting) and long distance movements (migration) \cite{Em_ref_1, Em_ref_2}. During the resting mode birds travel up to 50 km but return to a central place every day, whereas during the migration mode birds do not return to the central place and can travel several hundreds of kilometers per day. One point of interest is that whilst the resting periods can last months on end, the migrations may only last for a few days on end. See Figure \ref{bird_sample_path_map} for a section of a sample path centred around London.

\subsubsection{Identification of states.}
The bird tracking data were collected by the UvA-BiTS system \cite{Em_ref_1} and contains tracks gathered from 10 birds over the months July until January in the years 2012 and 2013. Approximately every few hours\footnote{Although some devices are configured differently to the extent that a recording is taken every $15$ minutes.}, a recording is taken of a global time-stamp along with the bird's current latitude and longitude coordinates. 

To identify the state of a given bird, we create a signal centred around a time point of interest which we threshold to determine whether the bird is either undergoing local or migratory behaviour. By considering all GPS coordinates in a 24 hour window, we calculate the diameter of the convex hull (or diameter of a minimum bounding circle)\footnote{The maximum distance between any two points in the set.} of the set by using the Haversine formula\footnote{The Haversine formula is an equation for great-circle distances between a pair of points on a sphere. By considering the radius of the sphere (i.e. the approximate radius of the Earth) and a pair of latitude--longitude co-ordinates, one can calculate the distance between them.}. This signal is sampled $10$ times a day. If the value of this signal is low, points are clustered together (local resting behaviour) otherwise they are spread apart (migratory behaviour). At the cost of including some erroneous exceptionally short rests, we can set a low threshold value of $52 \text{km}$; the presence of short rests is then fixed by discarding any resting phases shorter than 2 days. In comparison, the running periods can virtually be of any length as there have been instances of a bird flying exceptionally long distances over a week. 

\subsubsection{Results.}
\begin{figure}
  \centering
  \begin{overpic}[width=0.5\textwidth]{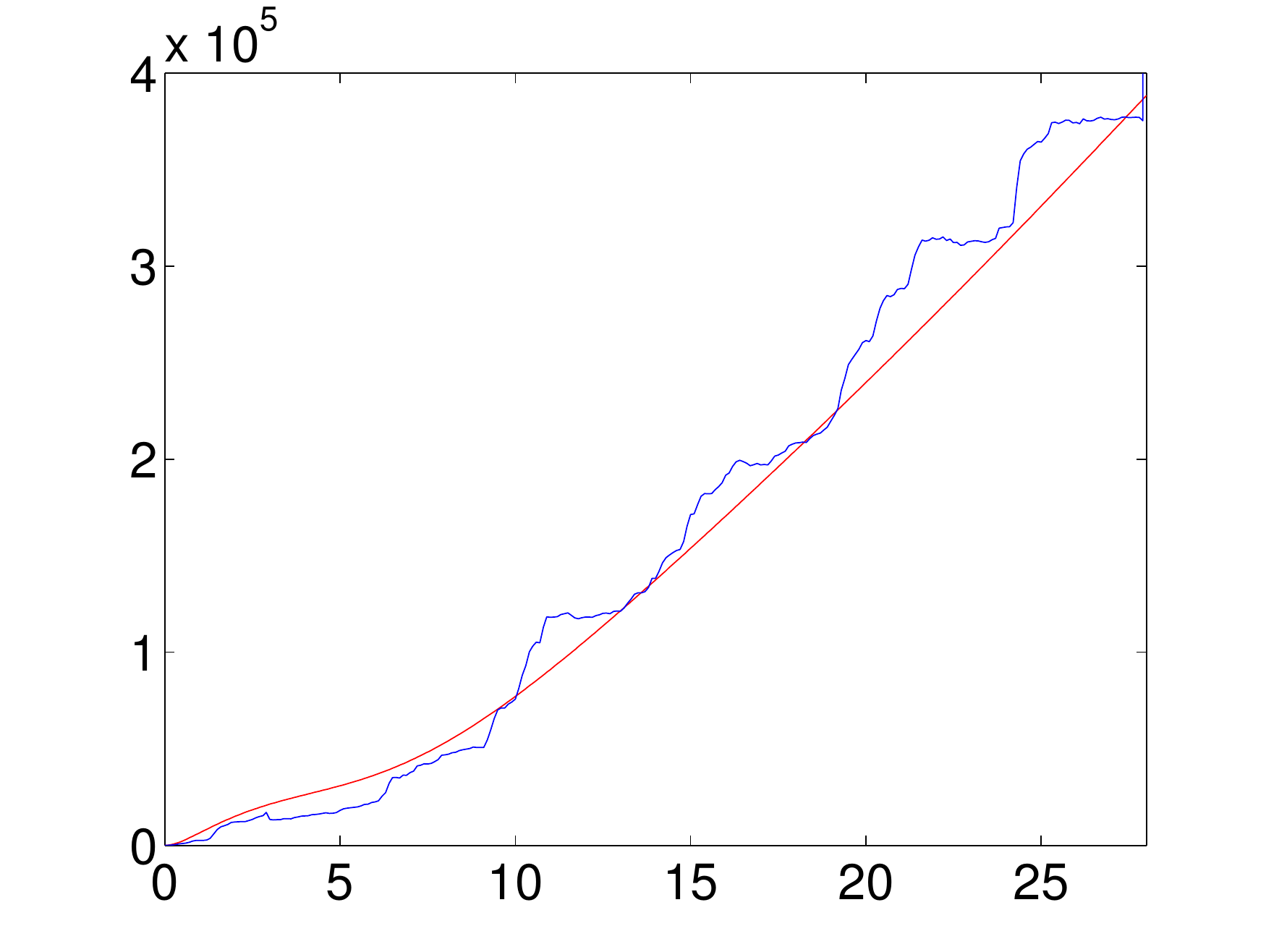}
  	\put(5,12){\rotatebox{90}{Mean sq. displacement $(\text{km})^2$}}
	\put(39,0){Time (days)}  
  \end{overpic}  
  \caption{\footnotesize{Comparison between system of equations (\ref{MSD_first_eq}--\ref{MSD_last_eq}) and \emph{Larus fuscus} data. In red, the theoretical value of $(D_p^2 + D_r^2)/N_0$ is plotted and in blue there is the experimentally derived average MSD calculated from the bird's initial position. From the data, it was extracted that $\tau \sim \text{IG}(1.26,1.22)$, $\omega \sim \text{IG}(10.79,7.42)$. Equally, for the system of differential equations, we specify $N_p(0) = 6$, $N_r(0) = 56$, $\psi_d = 0.42$ and $S_T^2 = 1.03\times 10^{5}\, (\text{km})^2 / \text{day}$. The initial state for all other differential equations is set to zero, except for $V_p^2(0) = S_T^2 N_p^2(0)$.}}
  \label{MSD_gulls}
\end{figure}

As we only had the data for $10$ birds available, we divided their sample paths up into $28$ day intervals after approximating distributions of interest, leading to calculation of the MSD over 62 sample paths. In contrast to the \emph{E. coli} dataset, we see that running and waiting times are non-exponentially distributed. The distribution of running and waiting times were approximated by Inverse Gaussian distributions $\tau \sim \text{IG}(1.26,1.22)$ and $\omega \sim \text{IG}(10.79,7.42)$. The speed distribution gave an estimate for the mean squared running speed as $S_T^2 = 1.03\times 10^{5}\, (\text{km})^2 / \text{day}$ and again using a Von Mises distribution in angle, we find $\psi_d = 0.42$.

In Figure \ref{MSD_gulls}, we plot the mean squared displacement in kilometres squared against time in days. As there were fewer sample paths available, the empirical mean square displacement curve is not very smooth and as a result the agreement with the theoretical curve is less good than in the bacterial case. However as the majority of the gulls were in a resting state to begin with, we do capture the initial delay before a linear growth stage. As the gulls are frequently resting as opposed to migrating, we see the data for the gulls (in blue) undergoing a style of step function where a small number of gulls undergoing fast movement quickly changes the MSD for the whole population. As the number of sample paths increases, this effect will smooth out.

With both examples, as time passes, we see that $D_p^2(t) + D_r^2(t) \sim t$ for large $t$. It is well known that linear mean squared displacement corresponds to the solution of the diffusion equation, or at least diffusive-like behaviour. It would now be pertinent to see if a diffusion approximation could be found for large time.

\section{Large time diffusion approximation.}

We now construct a large time effective diffusion equation. By first considering equations (\ref{p_forward})--(\ref{r_forward}), we transform into Laplace space, where large values of $t$ correspond to small values of the Laplace variable $\lambda$.  We then carry out a Taylor expansion of the delay kernels to remove the convolutions in time (see equations (\ref{p_laplace})--(\ref{r_laplace}) in Appendix \ref{AppA} for details).

Converting back to the time domain, one obtains
\begin{equation}\label{large_time_p_for}
(1 + \bar{\Phi}_\tau '(0))\left(\frac{\p}{\p t} + \vect{v}\cdot{\nabla_{\vect{x}}} \right) p = -\bar{\Phi}_\tau(0) p  + \int_V T(\vect{v},\vect{v}')\left( \bar{\Phi}_\omega(0) r(t,\vect{x},\vect{v}') + \bar{\Phi}_\omega'(0)\frac{\p}{\p t} r(t,\vect{x},\vect{v}')    \right) \dd\vect{v}' ,
\end{equation}
and
\begin{equation}\label{large_time_r_for}
(1 + \bar{\Phi}_\omega'(0))\frac{\p}{\p t}r =  -\bar{\Phi}_\omega(0) r  +\bar{\Phi}_\tau(0) p + \bar{\Phi}_\tau'(0)\left(\frac{\p}{\p t} + \vect{v}\cdot{\nabla_{\vect{x}}} \right) p 
\end{equation}

There are now two further steps to obtain an effective diffusion equation. First, by considering successively greater monomial moments in the velocity space, one obtains a system of $k$-equations where the equation for the time evolution of moment $k$ corresponds to the flux of moment $k+1$. It therefore becomes necessary to `close' the system of equations to create something mathematically tractable. We use the Cattaneo approximation for this purpose \cite{Hillen_2003, Hillen_2004}. Once a closed system of equations has been found, we then carry out an asymptotic expansion where we investigate the parabolic regime to obtain a single equation for the evolution of the density of particles at large time.

Note that it would be possible to carry out a similar process for smaller time behaviour by Taylor expanding the spatial delays in the convolution integrals. Asymptotic analysis would then have to be carried out to simplify the remaining convolution.

\subsection{Moment equations}
We can multiply equations (\ref{large_time_p_for})--(\ref{large_time_r_for}) by monomials in $\vect{v}$ and integrate over the velocity space to obtain equations for the velocity moments
\begin{equation}
m^0_\rho = \int_V \rho(t,\vect{x},\vect{v})\dd \vect{v}, \quad \vect{m}^1_\rho = \int_V \vect{v}\rho(t,\vect{x},\vect{v})\dd \vect{v}, \quad M^2_\rho = \int_V \vect{v}\vect{v}^T\rho(t,\vect{x},\vect{v})\dd \vect{v}.
\end{equation}
The equations relating the terms $m_p^0, m_r^0, \vect{m}_p^1, \vect{m}_r^1, M_p^2$ are given below. For initial integration over the velocity space, we see
\begin{eqnarray}\label{m0_p}
(1 + \bar{\Phi}_\tau'(0))\left(  \frac{\p m^0_p}{\p t} + \nabla_{\vect{x}} \cdot \vect{m}_p^1 \right) = - \bar{\Phi}_\tau(0)m_p^0 + \bar{\Phi}_\omega(0)m_r^0 + \bar{\Phi}_\omega'(0)\frac{\p m_r^0}{\p t}  , 
\end{eqnarray}
and
\begin{eqnarray}\label{m0_r}
(1 + \bar{\Phi}_\omega'(0))\frac{\p m^0_r}{\p t} = - \bar{\Phi}_\omega(0)m_r^0 + \bar{\Phi}_\tau(0)m_p^0 + \bar{\Phi}_\tau'(0)\left( \frac{\p m^0_p}{\p t} + \nabla_{\vect{x}} \cdot \vect{m}_p^1 \right),
\end{eqnarray}
When summing equations (\ref{m0_p}) and (\ref{m0_r}), we see that mass flux is caused by the movement of particles in the running state only, i.e. 
\[
\frac{\p }{\p t}\left( m_p^0 + m_r^0 \right)  + \nabla_{\vect{x}} \cdot \vect{m}_p^1 = 0.
\]
For multiplication by $\vect{v}$ and integrating, we obtain equations
\begin{eqnarray}
(1 + \bar{\Phi}_\tau'(0))\left(  \frac{\p \vect{m}^1_p}{\p t} + \nabla_{\vect{x}} \cdot  M_p^2 \right) = - \bar{\Phi}_\tau(0)\vect{m}_p^1 + \psi_d \bar{\Phi}_\omega(0)\vect{m}_r^1 + \psi_d \bar{\Phi}_\omega'(0)\frac{\p \vect{m}_r^1}{\p t}  ,  
\end{eqnarray}
and
\begin{eqnarray}
(1 + \bar{\Phi}_\omega'(0))\frac{\p \vect{m}^1_r}{\p t} = - \bar{\Phi}_\omega(0)\vect{m}_r^1 + \bar{\Phi}_\tau(0)\vect{m}_p^1 + \bar{\Phi}_\tau'(0)\left( \frac{\p \vect{m}^1_p}{\p t} + \nabla_{\vect{x}}\cdot  M_p^2 \right).
\end{eqnarray}
We would now like to approximate the $M_p^2$ term to close the system.
\subsection{Cattaneo approximation step}

We make use of the Cattaneo approximation to the velocity jump equation as studied by Hillen \cite{Hillen_2003, Hillen_2004}. For the case where the speed distribution is independent of the previous running step, i.e $h(s,s') = h(s)$, we approximate $M_p^2$ by the second moment of some function $u_\text{min} = u_\text{min} (t,\vect{x},\vect{v})$, such that $u_\text{min}$ has the same first two moments as $p = p(t, \vect{x}, \vect{v})$ and is minimised in the $L^2(V)$ norm weighted by $h(s)/s^{n-1}$. This is essentially minimising oscillations in the velocity space whilst simultaneously weighting down speeds which would be unlikely to occur \cite{Hillen_2003}.

We introduce Lagrangian multipliers $\Lambda^0 = \Lambda^0(t, \vect{x})$ and $\vect{\Lambda}^1 =\vect{\Lambda}^1(t, \vect{x})$ and then define
\begin{equation}
H(u) := \frac{1}{2} \int_V \frac{u^2}{h(s)/s^{n-1}}\dd \vect{v} - \Lambda^0\left(\int_V u \dd \vect{v} - m_p^0\right) - \vect{\Lambda}^1\cdot \left(\int_V\vect{v} u \dd \vect{v} - \vect{m}_p^1\right).
\end{equation}
By the Euler-Lagrange equation \cite{Gregory}, we can minimise $H(u)$ to find that
\begin{equation}
u(t, \vect{x}, \vect{v})  = \frac{\Lambda^0(t, \vect{x}) h(s)}{s^{n-1}} + \frac{(\vect{\Lambda}^1(t, \vect{x}) \cdot \vect{v}) h(s)}{s^{n-1}}.
\end{equation}
We now use the constraints to find $\Lambda^0$ and $\vect{\Lambda}^1$. For $m_p^0$ we have
\begin{equation}
m_p^0 = \int_V u \dd \vect{v} =  \Lambda^0 \int_V  h(s)/s^{n-1} \dd \vect{v}  = \Lambda^0 \text{Area}(\mathds{S}^{n-1}),
\end{equation}
where $\mathds{S}^{n} = \{\vect{x}\in\mathds{R}^{n+1}:\vectornorm{\vect{x}} = 1\}$ is the $n$-sphere centred at the origin. Notice also that the $\int_V \vect{v}h(s)/s^{n-1} \dd \vect{v} = \vect{0}$ by symmetry. For the first moment, we calculate
\begin{equation}
\vect{m}_p^1 = \int_V \vect{v} u \dd \vect{v} =   \vect{\Lambda}^1 \cdot \int_V  \vect{v}\vect{v}^T h(s)/s^{n-1} \dd \vect{v}  = S^2_T \text{Vol}(\mathds{V}^{n})\vect{\Lambda}^1,
\end{equation}
where $\mathds{V}^n$ is the closure of $\mathds{S}^{n-1}$, i.e. the ball around the origin. Therefore, we can stipulate the form for $u_{\text{min}}$ as
\begin{equation}
u_{\text{min}}(t, \vect{x}, \vect{v})  = \frac{m_p^0(t, \vect{x}) h(s)}{s^{n-1}\text{Area}(\mathds{S}^{n-1})} + \frac{(\vect{m}_p^1(t, \vect{x}) \cdot \vect{v}) h(s)}{S_T^2s^{n-1}\text{Vol}(\mathds{V}^n)}.
\end{equation}
We now approximate the second moment of $p$ by the second moment of $u_{\text{min}}$. 
\begin{equation}
M^2(u_{\text{min}}) = \int_V \vect{v}\vect{v}^T u_{\text{min}}(t,\vect{x},\vect{v})\dd \vect{v} = S_T^2 \frac{\text{Vol}(\mathds{V}^n)}{\text{Area}(\mathds{S}^{n-1})} I_n  m_p^0(t, \vect{x}) = \frac{S_T^2}{n}I_n m_p^0(t, \vect{x}) .
\end{equation}
So in the above equations, we simply approximate $\nabla_{\vect{x}} \cdot M_p^2 \approx \frac{S_T^2}{n}\nabla_{\vect{x}}m_p^0$.

\subsection{Effective diffusion constant}
Finally, we rescale our equations using the parabolic regime \cite{Erban_2004}
\begin{equation}
\begin{array}{ccc}
t = \hat{t}/\varepsilon^2,  &\quad\ & \vect{x} = \hat{\vect{x}} /\varepsilon ,
\end{array}
\end{equation}
for arbitrary small parameter $\varepsilon > 0$. By putting our variables into vectors $\vect{u} = (m_p^0, m_r^0)^T$ and $\vect{v}=(\vect{m}_p^1,\vect{m}_r^1)^T$, we drop the hats over the rescaled variables and rewrite our equations as
\begin{equation}
\varepsilon^2 \frac{\p}{\p t} A\vect{u} + \varepsilon F \nabla_{\vect{x}}\cdot  \vect{v} = C\vect{u}, \quad \varepsilon^2 \frac{\p}{\p t} B\vect{v} + \varepsilon \frac{S_T^2}{n}F \nabla_{\vect{x}} \vect{u} = D\vect{v},
\end{equation}
where $\nabla_{\vect{x}}  \vect{u} = [\nabla_{\vect{x}}m_p^0,\nabla_{\vect{x}}m_p^0]^T$ and $\nabla_{\vect{x}} \cdot \vect{v} = [\nabla_{\vect{x}}\cdot \vect{m}_p^1,\nabla_{\vect{x}}\cdot \vect{m}_p^1]^T$. Our time derivative matrices are given by
\begin{equation}
A = \left[
\begin{array}{cc}
1 + \bar{\Phi}_\tau '(0) & - \bar{\Phi}_\omega '(0)  \\
- \bar{\Phi}_\tau '(0) & 1 + \bar{\Phi}_\omega '(0)  \\
\end{array} \right], 
\quad
B = \left[
\begin{array}{cc}
1 + \bar{\Phi}_\tau '(0) & - \psi_d\bar{\Phi}_\omega '(0)  \\
- \bar{\Phi}_\tau '(0) & 1 + \bar{\Phi}_\omega '(0)  \\
\end{array}
\right],
\end{equation}
our flux matrix is given as
\begin{equation}
F = \left[
\begin{array}{cc}
 1 + \bar{\Phi}_\tau '(0) & 0 \\
 - \bar{\Phi}_\tau '(0) & 0 \\
\end{array}
\right] .
\end{equation}
Finally our source terms are
\begin{equation}
C = \left[
\begin{array}{cc}
- \bar{\Phi}_\tau (0) &  \bar{\Phi}_\omega (0) \\
  \bar{\Phi}_\tau (0) & -\bar{\Phi}_\omega (0) \\
\end{array}\right] \quad
D = \left[
\begin{array}{cc}
 - \bar{\Phi}_\tau (0) &  \psi_d \bar{\Phi}_\omega (0) \\
  \bar{\Phi}_\tau (0) & - \bar{\Phi}_\omega (0) \\
\end{array}\right]
\end{equation}
By using the regular asymptotic expansion
\begin{equation}
\vect{u} = \vect{u}^0 + \varepsilon \vect{u}^1 + \varepsilon^2 \vect{u}^2 + ... ,\quad \vect{v} = \vect{v}^0 + \varepsilon \vect{v}^1 + \varepsilon^2 \vect{v}^2 + ...
\end{equation} 
for $\vect{u}^j = (m_{p(j)}^0, m_{r(j)}^0)^T$ and $\vect{v}^j = (\vect{m}_{p(j)}^1,\vect{m}_{r(j)}^1)^T$, we obtain the set of equations
\begin{equation}
\begin{array}{ll}
{\varepsilon^0:} &  C\vect{u}^0 = \vect{0} ,\quad D\vect{v}^0 = \vect{0}, \\
{\varepsilon^1:} &   F\nabla_{\vect{x}}\cdot\vect{v}^0 = C\vect{u}^1 , \quad F\nabla_{\vect{x}}\vect{u}^0 = D\vect{v}^1,\\
{\varepsilon^2:} &  \frac{\p}{\p t} A\vect{u}^0 + F \nabla_{\vect{x}}\cdot  \vect{v}^1 = C\vect{u}^2 , \\
& \frac{\p}{\p t} B\vect{v}^0 +\frac{S_T^2}{n}F \nabla_{\vect{x}}  \vect{u}^1 = D\vect{v}^2 .
\end{array}
\end{equation}
Providing $\psi_d \not= 1$, solving these in order gives rise to the differential equation for total density $m^0 = m_{p(0)}^0 + m_{r(0)}^0$
\begin{equation}
\frac{\p }{\p t} m^0 = D \nabla_{\vect{x}}^2 m^0,
\end{equation}
for
\begin{equation}
D = \frac{S_T^2}{n}\frac{1}{\bar{\Phi}_\tau (0)}\frac{\bar{\Phi}_\omega (0)}{\bar{\Phi}_\omega (0) + \bar{\Phi}_\tau (0)} \frac{1 + \bar{\Phi}_\tau ' (0)(1-\psi_d)}{1 - \psi_d}.
\end{equation}
We now wish to find the values of $\bar{\Phi}_\tau (0), \bar{\Phi}_\omega (0)$ and $ \bar{\Phi}_\tau '(0)$. For probability distributions defined over the positive numbers with pdf $f(t)$, we see that the Laplace transform can be Taylor expanded as
\begin{equation}
\bar{f}(\lambda)  = 1 - \lambda \langle t \rangle +  \lambda^2 \langle t^2 \rangle - ...
\end{equation} 
for small $\lambda$. Therefore, by putting these terms into the expression $\bar{\Phi}(\lambda)$ given by equation (\ref{f_psi_conversion}), provided that all moments are finite, we see that
\begin{equation}
\bar{\Phi}_i (0)= \lim_{\lambda \to 0}\bar{\Phi}_i(\lambda)  = 1/\mu_i, \quad \bar{\Phi}_i '(0)= \lim_{\lambda \to 0}\bar{\Phi}_i'(\lambda) =  (\sigma_i^2 - 1)/2\mu_i^2,\quad\text{for }i=\tau , \omega ,
\end{equation}
for mean $\mu_i$ and variance $\sigma_i^2$ of distribution $i=\tau , \omega$, therefore
\begin{equation}
D = \frac{S_T^2}{n}\frac{\mu_\tau^2}{\mu_\tau + \mu_\omega} \left[ \frac{1}{1 - \psi_d} + \frac{1}{2}\left(\frac{\sigma_\tau^2}{\mu_\tau^2} - 1\right) \right] . 
\end{equation}
It is noteworthy that the variance of the running time distribution contributes to the diffusion constant, while it is independent of the variance of the waiting time distribution. Furthermore, when the running time distribution is exponentially distributed, the correction $\bar{\Phi}_\tau'(0)$ is identically zero. So we can view our diffusion constant as the contribution from the exponential component of the running time distribution, plus an additional term for non-exponential running times.

When referring back to the experimental data, it can be seen that by the end of the $4$ seconds, the \emph{E. coli} has entered into the diffusive regime with $D \approx 12.5\,(\mu\text{m})^2 / \text{s}$. The \emph{L. fuscus} however is yet to reach this state; we can predict that when it does, the corresponding value of the diffusion constant will be $D \approx 4.7\times 10^4\,(\text{km})^2 / \text{day}$, the solution of the mean squared displacement equations for greater time periods suggests that this is true.

\subsection{Numerical example.}
\begin{figure}[h!]
\centering
\includegraphics[width=0.30\textwidth]{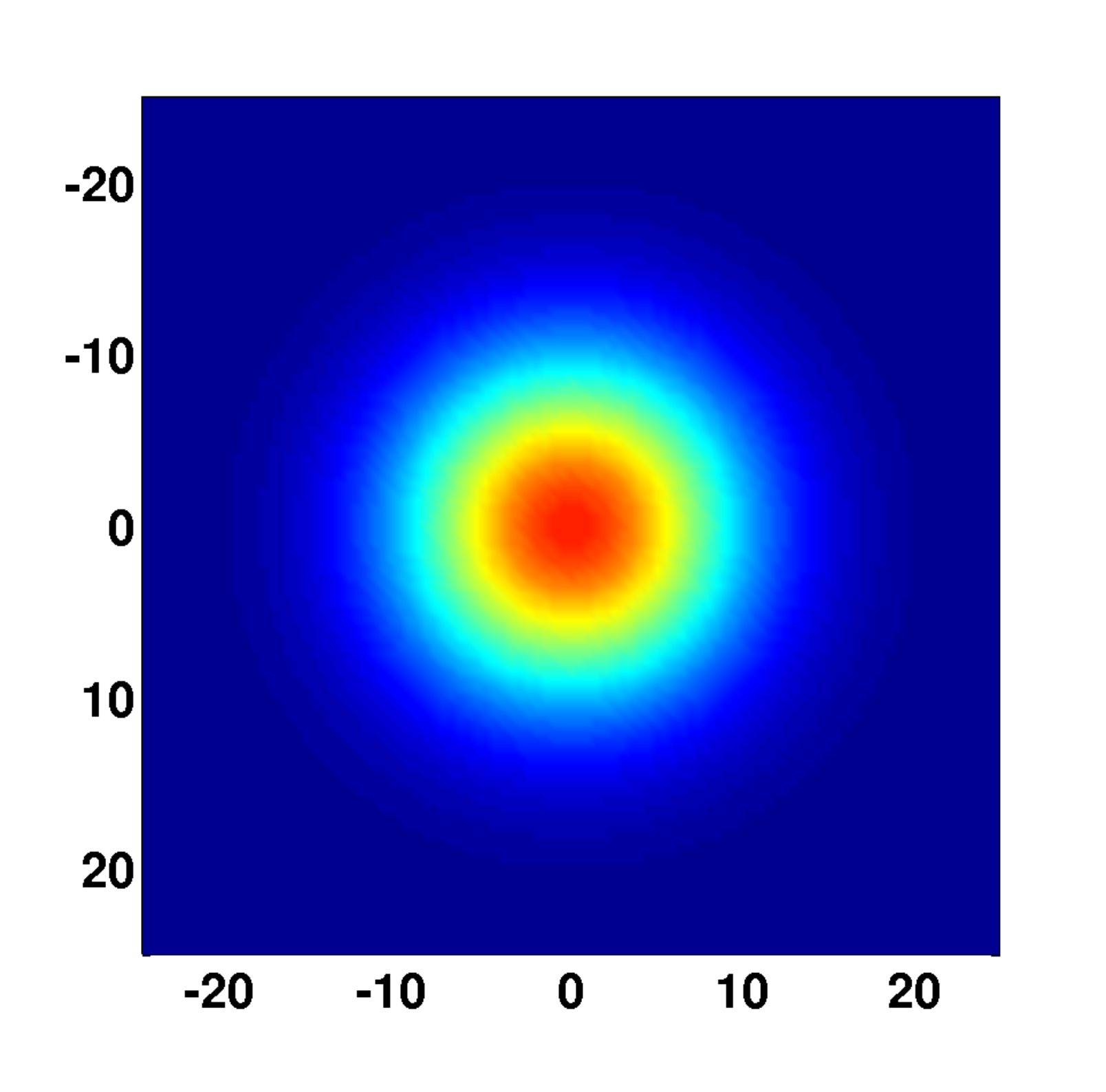}
\includegraphics[width=0.30\textwidth]{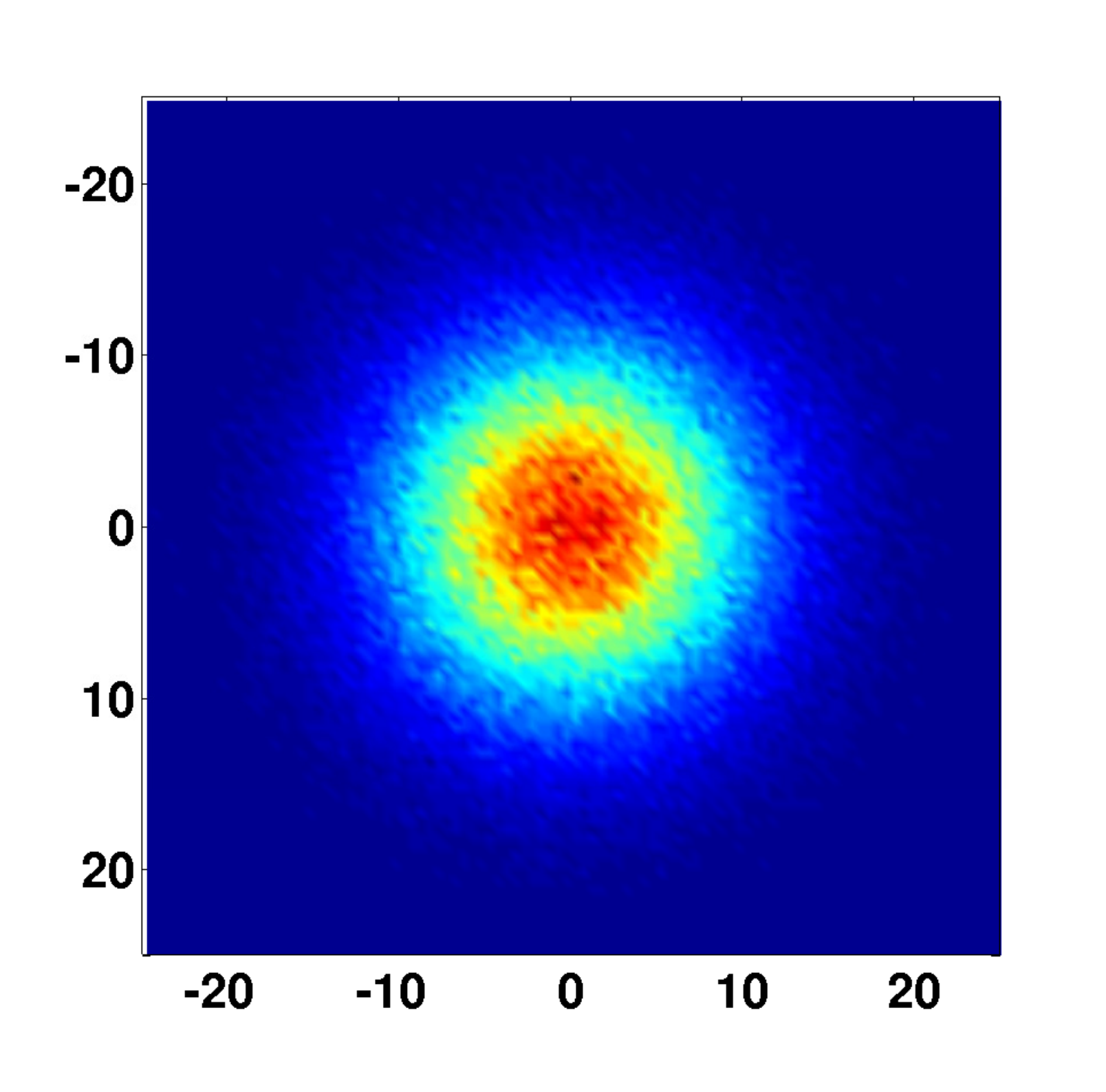}
\includegraphics[width=0.034\textwidth]{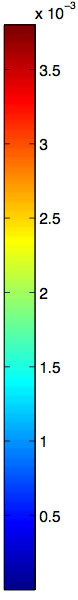}  
\\
\includegraphics[width=0.30\textwidth]{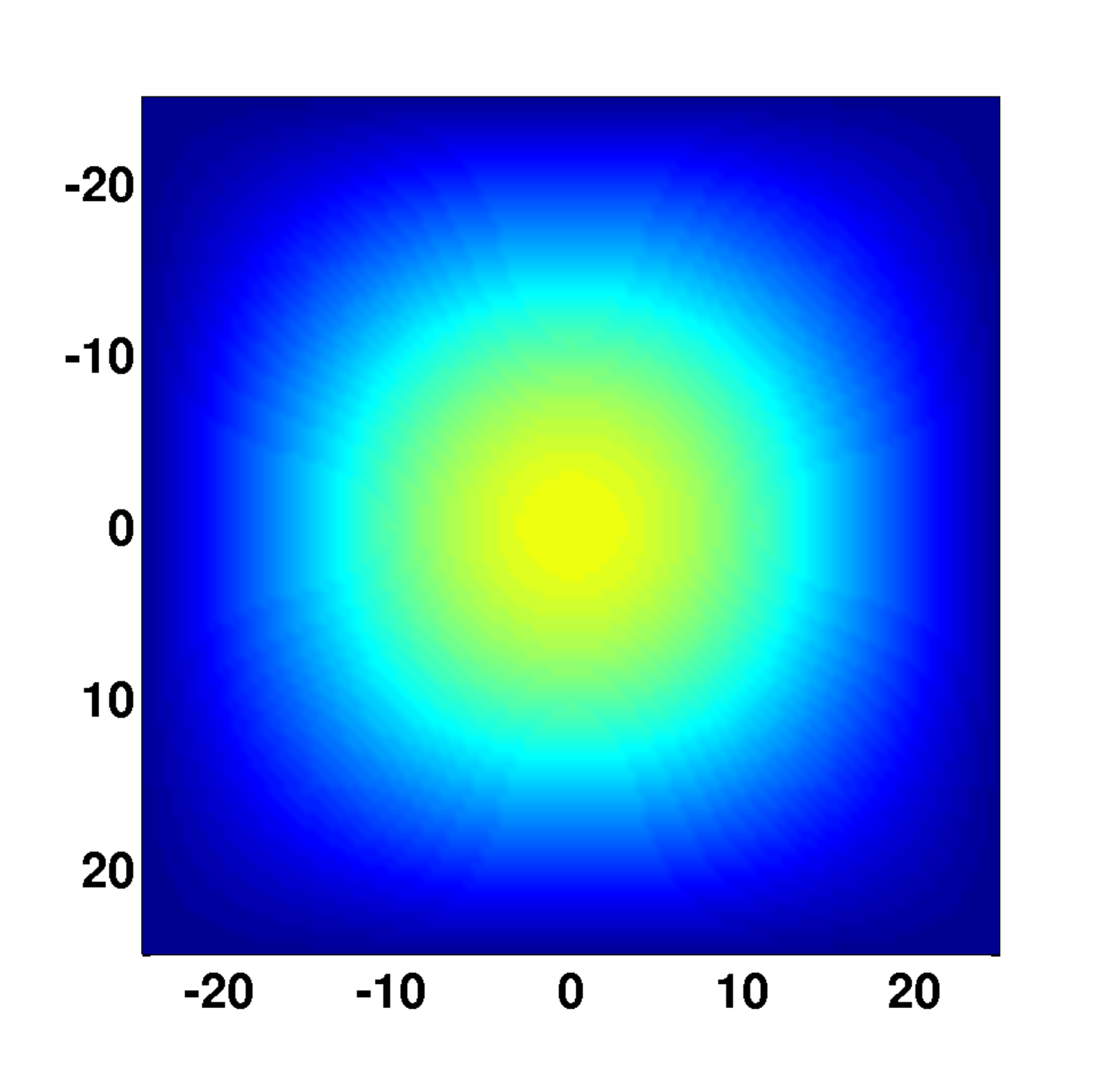}
\includegraphics[width=0.30\textwidth]{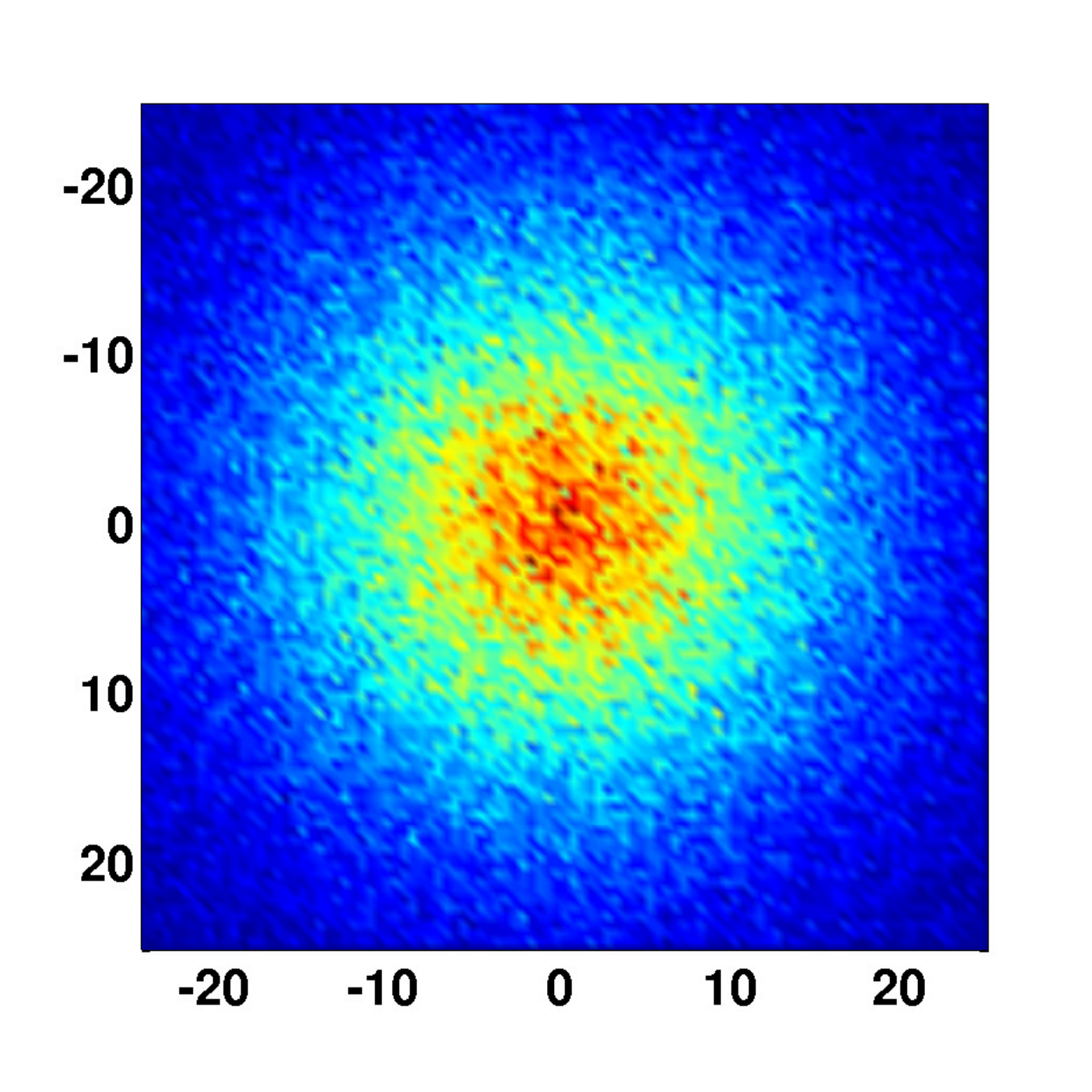}
\includegraphics[width=0.034\textwidth]{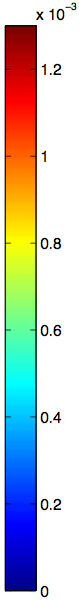} 
\caption{\footnotesize{Comparison between solution to the diffusion equation (\emph{left column}) and Gillespie simulation (\emph{right column}). \emph{Top row}: case where $\tau\sim\text{Exp}(1)$ and $\omega\sim\text{Exp}(1)$. \emph{Bottom row}: case where $\tau\sim\text{Gamma}(1/7, 7)$ and $\omega\sim\text{Gamma}(1/14,14)$.} For all Gillespie simulations, $3\times 10^5$ runs were carried out with half initialised in a running phase and half initialised in the resting phase.}
      \label{Comp_1}
\end{figure}

We now carry out a comparison between the underlying differential equation and Gillespie simulation. In Figure \ref{Comp_1}, we see the solution to the diffusion equation on the $\mathds{R}^2$ plane for a delta function initial condition, which takes the form of a bivariate Gaussian, compared with data simulated using  the algorithm given in Section \ref{TwoStateGVJ}. For the Gillespie simulation, all sample paths are initialised at the origin with fixed speed equal to unity and uniformly random orientation, half the sample paths are initialised in a run and half are initialised in a rest. Therefore all plots will have the parameters $S_T^2 = 1$, $\psi_d = 0$ and we specify $\mu_\tau = \mu_\omega = 1$, plots are shown at $t=100$.

On the top row, we see a diffusion approximation on the left compared with a velocity jump process where both the running and waiting times are sampled from an exponential distribution, with the mean of these distributions as stated, our effective diffusion constant for large time is $D = 1/2n$. On the bottom row, we see a diffusion approximation on the left compared with a velocity jump process where the running time is $\tau\sim\text{Gamma}(1/7,7)$ distributed, giving $\mu_\tau = 1$ and $\sigma_\tau^2 = 7$, the diffusion constant is therefore $D = 2/n$. The waiting time is $\omega\sim\text{Gamma}(1/14, 14)$ distributed; the high variance of the waiting time is chosen such that the simulation relaxes towards the diffusion approximation quickly. It was seen from numerical simulations that there there is a relationship between the choice of distribution for the waiting time $\omega$ and the index of persistence $\psi_d$ which will encourage the system to relax rapidly into the parabolic regime. It is necessary for the system to relax quickly in order for us to use the diffusion approximation as a valid method of comparison.

The choices of these two distributions was chosen to illustrate the importance of the diffusion correction term. This is illustrated in Figure \ref{Comp_1} by the difference between the top and bottom rows, which differ only in this correction term.

\begin{figure}[h!]
\centering
\includegraphics[width=0.30\textwidth]{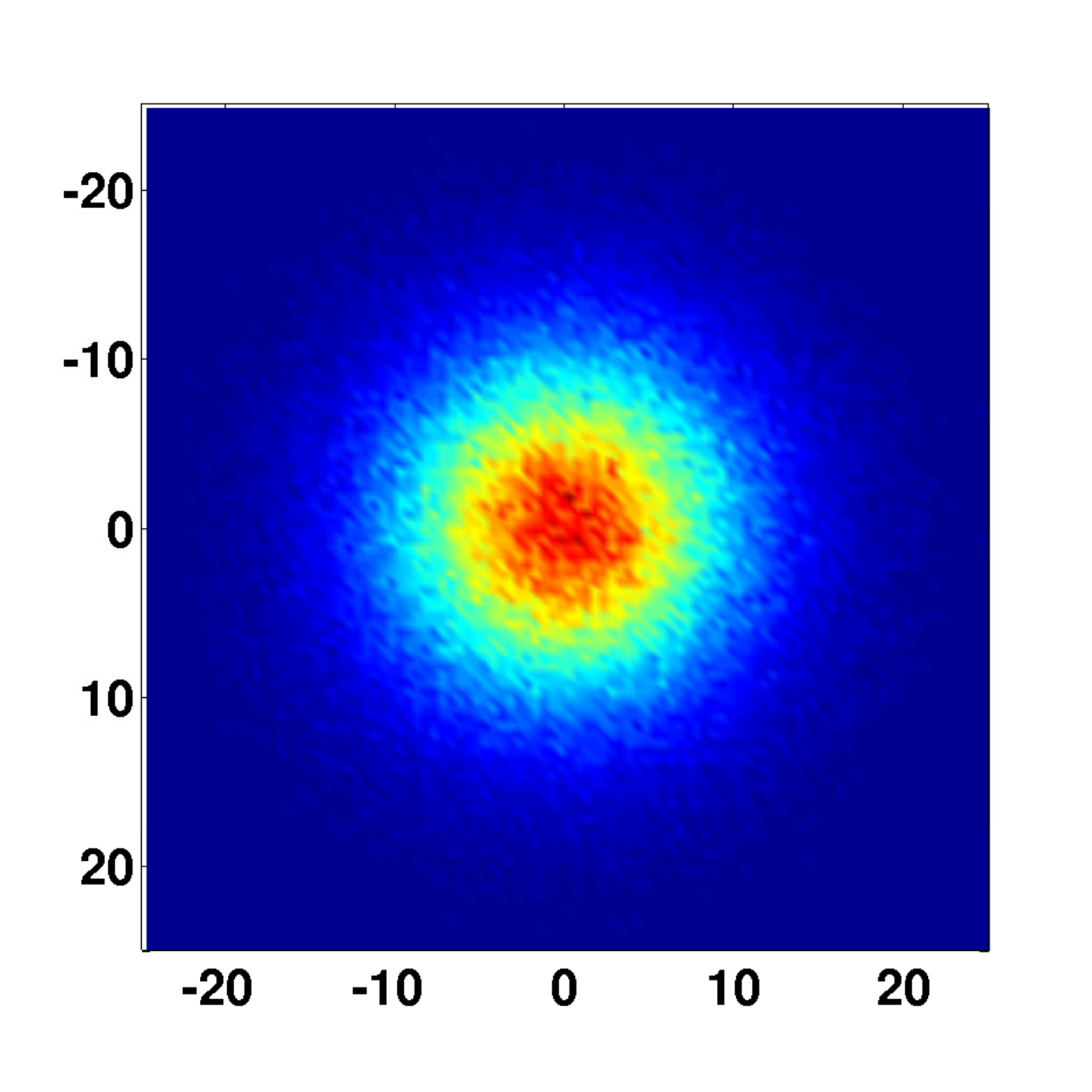}  
\caption{\footnotesize{Gillespie simulation for $\tau\sim\text{Gamma}(1/5, 5/2)$ and $\omega\sim\text{Exp}(1)$. For the Gillespie simulation, $2\times10^5$ simulations were initialised in a running phase and $10^5$ were initialised in a resting phase.}}
      \label{Comp_2}
\end{figure}

Another point of interest is that one can model distributions other than exponential with different means and still achieve the same effective diffusion constant through careful selection of variance. An example is shown in Figure \ref{Comp_2}, where the diffusion constant $D = 1/2n$ is recovered by changing the running distribution to $\tau\sim\text{Gamma}(1/5, 5/2)$. This then gives a mean run time of $\mu_\tau = 1/2$ and variance $\sigma_\tau^2 = 5/4$ and compares well to the first row in Figure \ref{Comp_1}. For this simulation, $2/3$ of the sample paths were initialised in a run and the remainder in a resting state so that the system was again encouraged to relax quickly.

\section{Final remarks.}

In this study, we have used a single modelling framework to describe two highly distinct biological movement processes, occurring in bacteria and birds.  In spite of the significant mechanistic differences between the two species, their phenomenological similarities nonetheless persist over length scales of 10 orders of magnitude. We recover the correct behaviour including the non-local delay effects due to non-exponential waiting times. This formulation could be considered a particularly phenomenological approach as it outlines a way for observables to directly parameterise movement equations. This is counter to some previous literature where quantities such as diffusion constants were left to the reader to identify \cite{McKenzie_2009}.  

A notable advantage of the modelling framework proposed here is the straightforward interpretation of the distributions and parameters involved, all of which have naturally intuitive meanings.  There is, unfortunately, no unified approach to extract such quantities of interest from biological movement data.  This was demonstrated in Section  \ref{CompExpTheo}, in which different approaches were taken to obtain the required parameters.  Nonetheless, such methods are the focus of much current research effort \cite{Gautestad_2013, Patterson_2008}, and we therefore believe that approaches such as ours will become increasingly relevant in the future. As far as the authors are aware, there has been no unified approach to tackling this problem.  

Finally, we demonstrated the novel result that for the underlying stochastic process of interest, the variance of the running time contributes to the large time diffusion constant. This raises the key question: when does the parabolic regime emerge? Our results also act as a warning against using the exponentially distributed running times as an approximation for other distributions, as whilst their mean values may align, the underlying dynamics can change drastically as shown with the link between Figures \ref{Comp_1} and \ref{Comp_2}.

Regarding the accuracy of this generalised velocity jump framework, it should be realised that the underlying models for the examples given could be improved by making the model more specific to the agent of interest. Below we discuss some possible alterations to the model - however at the cost of species generality.
\subsection{Extensions to model.}
For \emph{E. coli}, the bacterium is always subject to diffusion; in theory, this should add to its mean squared displacement while resting and may also affect running phases via rotational diffusion \cite{Rosser_2014}. If one wanted to incorporate a small fix to the resting state, it would be simple to add a diffusion term in space to equation (\ref{r_forward}). However, for a more comprehensive solution to the problem, to retain the correlation effects with turning kernel $T$, the equation (\ref{p_forward}) would have a rotational diffusion term added, which is achieved via a Laplacian in the velocity space \cite{Chandrasekhar_1943}. Furthermore, equation (\ref{r_forward}) would have to have to retain its defunct velocity field for orientation but also to include another velocity variable to allow for movement due to diffusion. 
A particularly interesting result would be to explore whether the Gaussian-like reorientation were a resultant effect from this rotational diffusion. By testing differing viscosities of fluid for the swimming \emph{E. coli}, one could undoubtedly make headway using this approach, work already initiated by Rosser \emph{et. al.} \cite{Rosser_2014}.

For the \emph{L. fuscus}, there are many physical and ecological phenomena which could to be built into the model; these range from the day-night cycles, in which the bird is reluctant to fly long distances through the night, to geographical effects, where the bird may follow the coastline for navigation. One could also consider environment factors, such as wind influence and availability of food resources. In the work by Chauviere \emph{et. al.} \cite{Chauviere_2010}, the authors consider the migration of cells along an extra-cellular matrix. Using a similar formulation to ours, but only considering exponentially distributed waiting times, cells are modelled to preferentially guide themselves along these extra-cellular fibres. It would not be difficult to imagine modifying this work to show how gulls may align their trajectory along coastlines, or using other geographical markers. 


\appendix

\section{Derivation of two-state Generalised Velocity Jump process.}\label{AppA}
We can motivate the set of equations (\ref{p_forward}-\ref{r_forward}) by considering the temporary variables:
\begin{itemize}
\item[$\eta$]$=\eta(t,\vect{x},\vect{v}) :=$ The density of particles at position $\vect{x}\,\dd \vect{x}$, with velocity $\vect{v}\,\dd \vect{v}$ at time $t\,\dd t$, \emph{having just started a jump}.
\item[$\nu$]$=\nu(t,\vect{x},\vect{v}):=$ The density of particles at position $\vect{x}\,\dd \vect{x}$, \emph{having just finished a jump} of velocity $\vect{v}\,\dd \vect{v}$ at time $t\,\dd t$ and \emph{just started a rest}. 
\end{itemize}
This leads to densities:
\begin{itemize}
\item[$p$]$=p(t,\vect{x},\vect{v}):=$ The density of particles at position $\vect{x}\,\dd \vect{x}$, with velocity $\vect{v}\,\dd \vect{v}$ at time $t\,\dd t$, being in a running state. We should note that we can relate $p$ to $\eta$ via the equation
\begin{equation}\label{p_in_eta}
p(t,\vect{x},\vect{v}) = \int_0^t F_\tau(t-s)\eta(s,\vect{x}-(t-s)\vect{v},\vect{v})\dd s = \int_0^t F_\tau(t-s)e^{-(t-s)\vect{v}\cdot \nabla_{\vect{x}}}\eta(s,\vect{x},\vect{v})\dd s,
\end{equation}
for $F_\tau(t) = \int_t^\infty f_\tau (s) \dd s$ being the probability that a jump lasts longer than $t$, clearly $F_\tau(0)=1$.  
\item[$r$]$=r(t,\vect{x},\vect{v}):=$ The density of particles at position $\vect{x}\,\dd \vect{x}$, having just finishing a jump of velocity $\vect{v}\,\dd \vect{v}$ at time $t\,\dd t$ in a resting state. Equally, there is the relation between $r$ and $\nu$, this is
\begin{equation}\label{r_in_nu}
r(t,\vect{x},\vect{v}) = \int_0^t F_\omega(t-s)\nu(s,\vect{x},\vect{v})\dd s, 
\end{equation}
for $F_\omega(t) = \int_t^\infty f_\omega (s) \dd s$ being the probability that a rest lasts longer than $t$, again $F_\omega(0)=1$.
\end{itemize}
By assuming that at time $t=0$, all particles are initiated into the beginning of a run with distribution $p_0(\vect{x},\vect{v})$, we can relate $\eta$ to previous times by the relationship
\begin{equation}\label{eta_in_nu}
\eta(t,\vect{x},\vect{v}) - p_0(\vect{x},\vect{v})\delta (t)= \int_V \int_0^t T(\vect{v},\vect{v}')f_\omega (t-s) \nu(s,\vect{x},\vect{v}') \dd s \dd \vect{v}'.
\end{equation}
Again by assuming that particles initiated into the beginning of a rest with distribution $r_0(\vect{x},\vect{v})$, there is the recursive relation for $\nu$
\begin{equation}\label{nu_in_eta}
\nu(t,\vect{x},\vect{v}) - r_0(\vect{x},\vect{v})\delta (t) = \int_0^t f_\tau (t-s) \eta(s,\vect{x}-(t-s)\vect{v},\vect{v}) \dd s = \int_0^t f_\tau (t-s) e^{-(t-s)\vect{v}\cdot \nabla_{\vect{x}}}\eta(s,\vect{x},\vect{v}) \dd s.
\end{equation}
Taking the Laplace transform in time of equations (\ref{p_in_eta}) and (\ref{r_in_nu}), we find
\begin{eqnarray}
\bar{p}(\lambda,  \vect{x}, \vect{v}) = \bar{F}_\tau(\lambda + \vect{v}\cdot \nabla_{\vect{x}}) \bar{\eta}(\lambda,  \vect{x}, \vect{v}), \\
\bar{r}(\lambda,  \vect{x}, \vect{v}) = \bar{F}_\omega(\lambda)\bar{\nu}(\lambda,  \vect{x}, \vect{v}).
\end{eqnarray}
Equally, taking the Laplace transform of (\ref{eta_in_nu}) and (\ref{nu_in_eta}), we see
\begin{eqnarray}
\bar{\eta}(\lambda,  \vect{x}, \vect{v}) - p_0(\vect{x},\vect{v}) = \int_V T(\vect{v},\vect{v}') \bar{f}_\omega (\lambda) \bar{\nu}(\lambda, \vect{x}, \vect{v}') \dd \vect{v}',\\
\bar{\nu}(\lambda,  \vect{x}, \vect{v}) - r_0(\vect{x},\vect{v}) = \bar{f}_\tau (\lambda + \vect{v}\cdot\nabla_{\vect{x}}) \bar{\eta}(\lambda, \vect{x}, \vect{v}).
\end{eqnarray}
Noting that in Laplace space
\begin{equation}
\bar{F}_i(\lambda) = 1-\bar{f}_i(\lambda)/\lambda  , \quad \text{for }i=\tau , \omega ,
\end{equation}
by eliminating $\eta$ and $\nu$, we derive paired differential equations in Laplace space
\begin{eqnarray}\label{p_laplace}
\left(\lambda + \vect{v}\cdot \nabla_{\vect{x}} \right)\bar{p}(\lambda, \vect{x},\vect{v}) - p_0(\vect{x},\vect{v}) = -\bar{\Phi}_\tau (\lambda + \vect{v}\cdot \nabla_{\vect{x}})\bar{p}(\lambda,\vect{x},\vect{v}) \\ +\bar{\Phi}_\omega(\lambda) \int_V T(\vect{v},\vect{v}')\bar{r}(\lambda, \vect{x},\vect{v}') \dd \vect{v}', \nonumber
\end{eqnarray}
and 
\begin{eqnarray}\label{r_laplace}
\lambda \bar{r}(\lambda, \vect{x},\vect{v}) - r_0(\vect{x},\vect{v}) = -\bar{\Phi}_\omega (\lambda)\bar{r}(\lambda,\vect{x},\vect{v}) +\bar{\Phi}_\tau(\lambda + \vect{v}\cdot \nabla_{\vect{x}}) \bar{p}(\lambda, \vect{x},\vect{v}),
\end{eqnarray}
where, as stated previously\footnote{We should be able to see the differential equation
\[
\frac{\dd F}{\dd t} = - \int_0^t \Phi(s)F(t-s) \dd s,
\]
must be satisfied by $\Phi$ - whether this is useful or not is another question!} 
\begin{equation}
\bar{\Phi}_i (\lambda) = \frac{\lambda \bar{f}_i (\lambda)}{1 - \bar{f}_i(\lambda)} = \frac{1 - \lambda \bar{F}_i(\lambda)}{\bar{F}_i(\lambda)},  \quad \text{for }i=\tau , \omega .
\end{equation}
Reverting back to the temporal variable $t$, we obtain equations (\ref{p_forward}--\ref{r_forward}).

\section{Properties of delay kernel $\Phi$.}\label{AppB}

The integro-differential equations for mean squared displacement, or indeed any other differential equation above, can now be easily solved by a variety of methods for numerical integration. However, in the case where one of the waiting times is exponentially distributed, $\Phi$ has been shown to become a multiple of the delta-function at the origin. It can be seen that many other distributions also have a numerical impulse at the origin, numerically integrating over an impulse is often difficult if not impossible so we carry out asymptotic analysis to evaluate the magnitude of said impulse.

To investigate the small time behaviour of $\Phi (t)$, we shall consider the small time behaviour of $f (t)$ then transform to Laplace space to consider large $\lambda$ behaviour and subsequently switch back.

By assuming the expansion of $f(t)$ to be of the form
\begin{equation}
f(t) \sim f_0 + f_1 t^{\alpha} + ... \quad \text{as }t\rightarrow 0.
\end{equation}
then subsequently in Laplace space
\begin{equation}
\bar{f}(\lambda) \sim \frac{f_0}{\lambda} + \frac{f_1 \Gamma (\alpha + 1)}{\lambda^{\alpha + 1}} + ... \quad \text{for } \alpha > -1,\text{ as }\lambda\rightarrow \infty .
\end{equation}
Using the relation (\ref{f_psi_conversion}) and considering the minimal contribution of the denominator, we find
\begin{equation}
\bar{\Phi}(\lambda) \sim f_0 + \frac{f_1 \Gamma (\alpha + 1)}{\lambda^\alpha} + ... \quad \text{for } \alpha > -1,\text{ as }\lambda\rightarrow \infty .
\end{equation} 
In the case where $\alpha >0$, we can subsequently invert to find
\begin{equation}
\Phi (t) \sim f_0 \delta (t) + f_1 \alpha t^{\alpha - 1} +  ... \quad \text{for } \alpha > 0,\text{ as } t \rightarrow 0 .
\end{equation}
From the above analysis, it should be clear we can expect an impulse at the origin for the case when $f_0 = f(0) \neq 0$ or $f_1\neq 0$ with $\alpha\in(0,1)$. By integrating between $0$ and $\varepsilon$, we see that
\begin{equation}\label{size_impulse}
\int_0^\varepsilon\Phi (t) \dd t \sim \int_0^\varepsilon \left[ f_0 \delta (t) + f_1 \alpha t^{\alpha - 1} +  ... \right] \dd t \sim f_0 + f_1 \varepsilon^\alpha \quad \text{for } \alpha > 0,\text{ as } \varepsilon \rightarrow 0 .
\end{equation}

\end{document}